Research paper

# Construction of Side Channel Attacks Resistant S-boxes using Genetic Algorithms based on Coordinate Functions

*B. Khadem\*[1], S. Rajav zade[2]*

1 Assistant Professor, Faculty of Computer Engineering, Imam Hossein Comprehensive University, Tehran.

2 Ph.D. student in Mathematics, Faculty of Mathematics, Payam-e-Noor University (PNU) Graduate Center, Tehran.

## Introduction

S-boxes are the main components of security in the nonlinear part of many cryptosystems, which are responsible for creating the property of confusion. Designing a robust cryptosystem usually requires the design of a resistant s-box. Hence, generating resistant s-box methods are among the topics of interest for cryptography researchers in the world.

In the last decade, numerous studies have been carried out on how to generate an appropriate s-box for cryptography applications. Overall, these construction methods can be divided into three categories: random methods, algebraic methods, and heuristic methods. Comprehensive search in the space of s-boxes [14], use of permutation algorithms [17], use of pseudo-random generators [29], use of key-dependent permutations [32], and use of pseudo-random generators with chaotic cores [10,19,29] are some typical methods in random category. The advantages of these methods include high-speed search, and its disadvantages include blind and even failed search.

Using the Bent functions [12] in order to withstand differential attacks and using the power mappings [8] such as the use of reverse power mapping in the AES-like s-box construction methods [11] are among the algebraic methods. Finding an algebraic approach that will undoubtedly lead to finding the ideal s-boxes yet remains an open problem.

However, finding targeted search techniques with time and memory complexity is possible in finite spaces. Therefore, heuristic methods such as the hybrid use of chaotic mapping and genetic algorithms [30] can result in finding suitable s-boxes by reducing time complexity. Some of these methods, such as those proposed in [18], employ one or more of the appropriate initial s-boxes in a random search. Others, like method [27], generate a set of equivalent s-boxes after the random construction of a suitable s-box. Among the known equivalences in Boolean vector functions, we can mention Carlet-Charpin-Zinoviev equivalence [2] and Extended Affine Equivalence [3], which are based on the displacement of coordinate functions of the initial s-box [1]. The authors in [20, 21] showed that the extended affine equivalence can be exploited to generate s-boxes with different transparency order (TO).

The construction method proposed in this paper consists of two phases. In the innovative pre-processing phase, a suitable initial s-box is generated. In the main phase, new boxes are produced using the linear combinations of the coordinate functions of the initial s-box generated in the preprocessing stage. The new boxes have been significantly improved in the properties of signal to noise rate (SNR), TO and CC.

This article's innovation is to employ a genetic algorithm on a linear combination set of coordinate functions of the initial s-box so that s-boxes resistant to side channel attacks (SCAs) are produced intelligently.

Subsequently, in Section 2, we introduce the basic concepts required for a s-box. In Section 3, we propose the method for the construction of s-boxes resistant to SCAs so that it does not decrease other properties. Section 4 expresses the results of the tests performed and presents some boxes produced by this method in the sizes of 4×4 to 8×8 and compares them with other known s-boxes. Eventually, in Section 5, we present a general conclusion and some future suggestions.

**Preliminary Definitions**

In this section, some of the concepts and definitions needed to study the rest of the article are provided. A vectorial Boolean function is a mapping in the form of $S: F_2^n \to F_2^m$, with the assumption of $m > 1$ as follows:

$$S(x_1, \dots, x_n) = (f_1(x_1, \dots, x_n), \dots, f_m(x_1, \dots, x_n)) \quad (1)$$

It is usually called an s-box and sometimes a (n,m)-function. In case m=1, s-box S is a Boolean function [31].

Functions $f_i: \{0,1\}^n \to \{0,1\}$ are Boolean n-variable functions called S coordinate functions. The Walsh transform of S is represented by $W_S$, which is defined on the field $F_2^n \times F_2^m$ as follows [26]:

$$W_S(w, u) = \sum_{0 \neq u \in F_2^m, w \in F_2^n} (-1)^{u.S(x) \oplus w.x} \quad (2)$$

In which $u.x = u_1.x_1 \oplus \dots \oplus u_n.x_n$ is the inner product on the field. We represent all linear combinations of coordinate functions of S, with the set

$$G = \{g_j: g_j = a.S, a \in F_2^n\} \quad (3)$$

In this paper Bijection, fixed points, and opposite fixed points are called primary properties. Nonlinearity degree (NL), Algebraic degree (AD), Correlation immunity (CI), differential uniformity (DU), Robustness to differential cryptanalysis (RDC), Algebraic immunity (AI), Absolute indicator, and Sum-of-square indicator are also called secondary properties. We also call the TO, SNR, and CC the properties for resistance to SCAs.

*A. Security Properties for the Evaluation of a S-Box*

The security properties of a s-box are indicators to represent their resistance against various attacks. For example, the properties of NL and CI indicate the resistance to linear attacks. The properties of RDC and DU against differential attacks, and the properties of FP and OFP represent resistance to algebraic attacks. Improving properties such as TO, SNR, and CC also cause the resistance against SCAs. Then, some of the most important of these properties will be introduced. $u \in F_2^n$, $v \in F_2^m$, and $i$ is an n-bit string. Multiplication and addition operators are also defined on the field.

*A.1. Balanced-ness*

We call S is balanced if $W_s(0, u) = 0$ [8]. The higher the magnitude of S-box imbalance, the high the probability of linear approximation is obtained.

*A.2. Nonlinearity degree*

NL is defined as follows [6]:

$$NL(S) = 2^{n-1} - \frac{1}{2} \max_{0 \neq v \in F_2^m, w \in F_2^n} \left| \sum_{x \in F_2^n} (-1)^{v.F(x) \oplus w.x} \right| \quad (4)$$

The high nonlinearity degree of an s-box increases the resistance against linear cryptanalysis.

*A.3. Algebraic Degree*

S can be uniquely expressed by the following polynomial:

$$S(x_1, \ldots, x_n) = \sum_{I \subseteq \{1,\ldots,n\}, a_I \in F_2^q} x^I a_I \quad (5)$$

In which $x^I = \prod_{i \in I} x_i$. As a result, the AD of S represented by deg (S) is defined as follows [24]:

$$deg(S) = max\{\#I | a_I \neq 0\}. \quad (6)$$

The low degree of s-box decreases the resistance against cryptanalytic attacks, namely algebraic attack, higher-order differential, interpolation, cube attacks [24].

*A.4. Correlation Immunity*

S is immune from the correlation of degree t if we have $W_s(w, u) = 0$ for every u, and every $w$ that $0 \leq hw(w) \leq t$ [9]. A s-box which is both balanced and correlation immune of $t^{th}$-order, is known as t-resilient. The high order resilient s-boxes are useful against the correlation attack [28].

*A.5. Differential Uniformity*

S is called $\delta$- uniform if and only if we have for each

$$0 \neq a \in F_2^n, z \in F_2^m$$

$$|\{x \in F_2^n; S(x) \oplus S(x \oplus a) = z\}| \in \{0, 1, 2, \ldots, \delta\} \quad (7)$$

Smaller values of $\delta$ are more appropriate [5]. The smaller value of Differential Uniformity increases the s-box resistance against differential cryptanalysis.

*A.6. Fixed Points and Opposite Fixed Points*

$x$ is a fixed-point of S if $S(x) = x$, and $x$ is an opposite fixed-point of S if $S(x) = x + 1$ [16]. We denote the number of fixed points of S by FP(S), and the number of opposite fixed points by OFP(S). The number of FP and OFP should be kept as low as possible to avoid leakage in any statistical cryptanalysis.

*A.7. Robustness against Differential Cryptanalysis*

If $L$ is the largest value of the differential characteristic table of S and $R$ is the number of non-zero values of it's first column (the value of $2^n$ is not calculated in the first row), then S is called $\varepsilon$-resistant to differential cryptanalysis and the value of $\varepsilon$ is expressed as follows [23]:

$$\varepsilon = \left(1 - \frac{R}{2^n}\right)\left(1 - \frac{L}{2^n}\right) \quad (8)$$

*A.8. Sum-of-Square Indicator*

If $\sigma$ is the symbol of the absolute indicator or the maximum value of absolute autocorrelation of S, which is derived from the autocorrelation function $\hat{r}_S$ then [25]:

$$\sigma = \max_{0 \neq v \in F_2^m} \sum_x \widehat{r_{v.S}}^2(\alpha) \quad (9)$$

*A.9. Absolute Indicator*

If $|AC(S)|_{max}$ is the symbol of the absolute indicator or the maximum value of absolute autocorrelation of S, which is derived from the autocorrelation function [25]

$$\hat{r}_S(\alpha, v) = \hat{r}_{v.S}(\alpha) = \sum_x (-1)^{v.S(x) \oplus v.S(x \oplus \alpha)} \quad (10)$$

Then

$$|AC(S)| = \max_{0 \neq v \in F_2^m, \alpha \in F_2^n} |\hat{r}_{v.S}(\alpha)|_{max} \quad (11)$$

with the low AC and $\sigma$, It is harder to perform an analysis of ciphertext, when trying to come up with an attack [33].

*A.10. Algebraic Immunity*

If $B_n$ is a set of n-variable Boolean functions, then the AI of S is equal to [7]:

$$AI_S = min\{deg(g): 0 \neq g \in B_n, \exists b \in F_2^m \ s.t. \ g|_{S^{-1}(b)} = 0\} \quad (12)$$

The large AI increases the resistance against algebraic attacks.

*A.11. Signal to Noise Ratio*

The SNR is defined as follows [15]:

$$SNR(DPA)(S) = m \, 2^{2n} \left(\sum_k \left(\sum_{i=0}^{n-1} \widehat{(-1)^{S_i}}(k)\right)^4\right)^{-1/2} \quad (13)$$

In which $\hat{f}(k) = \sum_x (-1)^{x.k} f(x)$ is the Walsh-Hadamard transform (WHT) of a Boolean function f. Increasing the SNR value of the s-box enhances its robustness against differential cryptanalysis, and decreasing it increases the robustness of the s-box to SCAs [15].

*A.12. Transparency Order*

The definition of TO represented as follows [26]:

$$TO_S = \max_{\beta \in F_2^n} \left(|m - 2hw(\beta)| - \frac{SS(a,\beta,v)}{2^{2n}-2^n}\right) \quad (14)$$

Where

$$SS(a,\beta,v) = \sum_{0 \neq a \in F_2^n} \left| \sum_{v \in F_2^m, hw(v)=1} (-1)^{v.\beta} W_{D_aS}(0,v) \right|$$

In which $W_{D_aS}$ is the Walsh-Hadamard transform of derivative S related to the vector $a \in F_2^n$. By reducing the amount of TO, the resistance of the s-box against relevant attacks such as DPA increases [23].

*A.13. Confusion Coefficient*

Suppose the selected function for DPA is a bit in the last round of encryption $\psi$ that depends on the encrypted cipher text bits, the subkey, and the s-box response. Therefore, if $(\psi|k_i)$ and $(\psi|k_j)$ are related to the assumed keys $\psi$ and $k_j$, we define the CC $\kappa$ on the two keys $(k_i, k_j)$ as follows [13]:

$$\kappa = \kappa(k_i, k_j) = Pr[(\psi|k_i) \neq (\psi|k_j)] = \frac{N_{(\psi|k_i) \neq (\psi|k_j)}}{N_t} \quad (15)$$

In which $N_t$ is the total number of values of the corresponding encrypted bits, and $N_{(\psi|k_i) \neq (\psi|k_j)}$ is the number of iterations that the keys $k_i$ and $k_j$ in different values $\psi$ generate different responses. Decreasing the value of CC increases the resistance against SCAs [25].

**The Proposed S-box Construction Method**

SCAs are among the most critical attacks on s-boxes, and decreasing the values of properties such as TO, SNR, and CC will increase the resistance against these attacks [15, 23, 25]. In this article, a proposed method is designed, intending to improve the resistance of generated s-boxes against SCAs without decreasing the resistance against other attack. This approach enhances this resistance by using a genetic algorithm on linear combinations of coordinate functions. Usually, s-boxes are evaluated from two perspectives of security and performance efficiency. The time and memory complexity required by the algorithm is called performance properties. The security properties of the s-box represent the degree of robustness to various attacks. It should be noted that some security properties have an inverse relationship with each other, meaning that improving one weakens the other. For example, AD and CI are in this way [28].

The main question of this study is whether it is possible to generate a new s-box by selecting and replacing some Boolean functions from the linear combinations of coordinate functions of an initial s-box, which, while maintaining the primary and secondary properties, improves the properties of robustness to SCAs.

Following the answer to the research question, it was initially found by theoretical investigations that there was no change in the set G by substituting $g_j$ instead of $f_i$ in S, and therefore the secondary properties would remain unchanged. On the other hand, because the value table of the coordinate functions changes, there will be a possibility of changing the primary properties and the properties for resistance to SCAs. Besides, in the preprocessing stage of the proposed method, evaluation and comparison have been done to select the appropriate s-box, sometimes with the production of 2 million s-boxes by random methods, methods of producing s-boxes in the Sage tool, and some other novel methods.

This scheme consists of two stages. The first stage is pre-processing, and the second stage is the main section. In the preprocessing stage, a suitable s-box is generated using a heuristic algorithm. The basic properties of this s-box are compared to the basic properties of known s-boxes. If their values are better or equal, the generated s-box is selected. This s-box is called S, and we employ it as the initial s-box in the main section.

The main section can be implemented as a genetic algorithm with the input of the s-box S from the preprocessing stage, which is as follows:

Stage 1. Extraction of coordinate functions of the s-box $S_{n \times n}$, i.e., $f_i$:

$$S(x_1, \ldots, x_n) = (f_1(x_1, \ldots, x_n), \ldots, f_n(x_1, \ldots, x_n)) \quad (16)$$

Stage 2. Generation of a set of all linear combinations of the coordinate functions S with the name G in expression (3), i.e., each of $g_j$ is as follows:

$$g_j = \sum_{i=0}^{n-1} a_i \cdot f_i, a_i \in F_2 \quad (17)$$

Thus, we will have $\#G = 2^n$.

Stage 3. n-random selection of $g_j$ and placing them as coordinate functions in a new s-box named $S_i$.

$$S_i = (g_0(x), \ldots, g_{n-1}(x)) \quad (18)$$

The order does not matter in this selection. On the basis of the observations made, the values of the s-box properties resulting from the displacement of the coordinate functions of an initial s-box are equal to the initial s-box in all properties except possibly fixed points and opposite fixed points.

After generating the s-boxes, there will be the possibility of changing the order of their coordinate functions.

Stage 4. Repetition of step 3 and generation of set C (number of members of set C is optional):

$$C = \{S_i : S_i(x) = (g_0(x), \ldots, g_{n-1}(x)), g_j \in G, x \in F_2^n\} \quad (19)$$

Therefore, we will have $\#C = \binom{2^n}{n}$.

Stage 5. Run of the bijective evaluation function on the members of set C and production of set B:

$$B = \{S_i : S_i \text{ is bijective}\} \quad (20)$$

At this stage, non-bijective s-boxes are removed.

Stage 6. Run of the function of fixed points and opposite fixed points on the members of set B and production of set F:

$$F = \{S_i : S_i \in B, FP(S_i) = 0, \ OFP(S_i) = 0\}. \quad (21)$$

At this point, s-boxes with fixed points or opposite fixed points are deleted.

Stage 7. Run of TO evaluation function on members of set F and production of set T:

$$T = \{S_i : S_i \in F, TO_{S_i} \geq TO_S\} \quad (22)$$

Therefore, the condition for the non-empty set of T is to find at least one s-box produced from the s-boxes, which is not firstly an initial s-box, secondly, it is bijective, has no fixed points, and TO is less than or equal to $TO_S$.

The above steps can be briefly illustrated in Algorithm 1. A short Python code implemented for the algorithm.

**The Results Achieved from the Proposed Scheme**

In this section, some s-boxes with dimensions of 4×4, 5×5, 6×6, 7×7, and 8×8 obtained from the proposed design are indicated in Tables 7 to 15, and the values of the properties gained from them are shown in Tables 2 to 6. The s-boxes generated are compared with the appropriate s-box and two other well-known s-boxes in these tables.

Algorithm1. Proposed Method Algorithm

**Input**: The initial S-box $S_{n \times n}$.
**Output**: The improved S-boxes set T.
1. Compute the set
2. $g = \{g_j : g_j = a.S, 0 \neq a \in F_2^n\}$.
3. **for** $i = 0$ to $\binom{2^n - 1}{n} - 1$ **do**
4. $S_i = (g_{j_0}, \ldots, g_{j_{n-1}}), j_k \in I$
5. **If** $FP(S_i) + OFP(S_i) + TO(S_i) = TO(S) \leq TO(S)$ **then**
6. Insert $S_i$ into the set T
7. **else**
8. break
9. **end for**
10. **until** reaching to maximum iteration allowed
11. **return** T

**Conclusion**

In this article, an approach is presented in which the properties related to SCAs such as SNR, TO, and CC are improved by using a suitable s-box, without reducing other security properties. Besides, some results obtained from examining s-boxes in the dimensions of 4×4, 5×5, 6×6, 7×7, and 8×8 demonstrated that the s-boxes generated are not only improved relative to the initial s-box, but in some cases, considerably better than the well-known s-boxes. On the basis of the observations made, the properties of TO and CC have a direct relationship with each other, and both have an inverse relationship with SNR. This means that the SNR decreases by improving the properties of TO and CC and vice versa so that a s-box that was better than the initial s-box in all three cases was not found. Along with related words, overall, the s-boxes generated are improved in all properties or have exactly the same values as the initial s-box properties. Some of the achieved results can be seen briefly in Table 1. Subsequently, it is proposed that the parallelism in the genetic algorithm employed to be taken into account so that s-boxes with dimensions larger than 8 × 8 can also be produced.

In Table 1, the symbols $\#T_{total}$, $\#T_{bij}$, $\#T_{FP}$, $\#T_{OFP}$, $\#T_{SNR}$, $\#T_{TO}$ and $\#T_K$ show the number of total generated $S_i$, the number of bijective $S_i$, the number of $S_i$ with $FP(S) = 0$, the number of $S_i$ with $OFP(S) = 0$, the number of $S_i$ with better SNR, the number of $S_i$ with better TO, and the number of $S_i$ with better $\kappa$, respectively.

Table 1: The number of generated s-boxes in different sizes

| Samples | 4x4 | 5x5 | 6x6 | 7x7 | 8x8 |
|---|---|---|---|---|---|
| $\#T_{total}$ | 1820 | 201376 | $10^5$ | $10^5$ | $10^4$ |
| $\#T_{bij}$ | 840 | 8332 | 40292 | 38185 | 4359 |
| $\#T_{FP}$ | 356 | 29399 | 13709 | 12978 | 1399 |
| $\#T_{OFP}$ | 339 | 36875 | 14934 | 16531 | 1207 |
| $\#T_{SNR}$ | 0 | 60 | 6 | 23 | 2413 |
| $\#T_{TO}$ | 355 | 310 | 9571 | 359 | 4273 |
| $\#T_K$ | 835 | 83238 | 40286 | 38164 | 1905 |
| $\#T_{better}$ | 70 | 220 | 2071 | 43 | 406 |

Table 2: Comparing the evaluation properties of 4×4 s-boxes

| Properties | 4x4 | | | initial | proposed | PRINCE | PRESENT |
|---|---|---|---|---|---|---|---|
| | min | avg | max | | | | |
| B | √ | √ | √ | √ | √ | √ | √ |
| NL | 4 | 4 | 4 | 4 | 4 | 4 | 4 |
| AD | 3 | 3 | 3 | 3 | 3 | 3 | 3 |
| CI | 0 | 0 | 0 | 0 | 0 | 0 | 0 |
| R | 0.75 | 0.75 | 0.75 | 0.75 | 0.75 | 0.75 | 0.75 |
| δ | 4 | 4 | 4 | 4 | 4 | 4 | 4 |
| AC | 8 | 8 | 8 | 8 | 8 | 8 | 8 |
| σ | 640 | 640 | 640 | 640 | 640 | 640 | 640 |
| AI | 2 | 2 | 2 | 2 | 2 | 2 | 2 |
| FP | 0 | 0.889 | 5 | 0 | 0 | 0 | 0 |
| OFP | 0 | 0.894 | 5 | 0 | 0 | 0 | 0 |
| SNR | 1.612 | 3.466 | 3.108 | 1.612 | 1.663 | 2.128 | 2.128 |
| TO | 3.4 | 3.539 | 3.733 | 3.533 | 3.466 | 3.4 | 3.533 |
| K | 0.157 | 0.457 | 1.357 | 1.357 | 1.357 | 0.657 | 0.657 |

Table 3: Comparing the evaluation properties of 5×5 s-boxes

| Properties | 5x5 | | | initial | proposed | Keecak | Primate |
|---|---|---|---|---|---|---|---|
| | min | avg | max | | | | |
| B | √ | √ | √ | √ | √ | √ | √ |
| NL | 10 | 10 | 10 | 10 | 10 | 10 | 10 |
| AD | 4 | 4 | 4 | 4 | 4 | 4 | 4 |
| CI | 0 | 0 | 0 | 0 | 0 | 0 | 0 |
| R | 0.937 | 0.937 | 0.937 | 0.937 | 0.937 | 0.937 | 0.937 |
| δ | 2 | 2 | 2 | 2 | 2 | 8 | 2 |
| AC | 8 | 8 | 8 | 8 | 8 | 8 | 8 |
| σ | 2048 | 2048 | 2048 | 2048 | 2048 | 2048 | 2048 |
| AI | 3 | 3 | 3 | 3 | 3 | 3 | 3 |
| FP | 0 | 1.029 | 7 | 0 | 0 | 1 | 1 |
| OFP | 0 | 0.81 | 7 | 0 | 0 | 1 | 1 |
| SNR | 2.307 | 3.432 | 4.603 | 2.361 | 2.517 | 3.922 | 3.690 |
| TO | 4.596 | 4.658 | 4.758 | 4.612 | 4.596 | 4.516 | 4.661 |
| K | 0.096 | 0.356 | 1.003 | 0.949 | 0.81 | 0.211 | 0.265 |

Table 4: Comparing the evaluation properties of 6×6 s-boxes

| Properties | 6x6 | | | initial | proposed | CTTL | Fides |
|---|---|---|---|---|---|---|---|
| | min | avg | max | | | | |
| B | √ | √ | √ | √ | √ | √ | √ |
| NL | 24 | 24 | 24 | 24 | 24 | 20 | 24 |
| AD | 5 | 5 | 5 | 5 | 5 | 5 | 5 |
| CI | 0 | 0 | 0 | 0 | 0 | 0 | 0 |
| R | 0.937 | 0.937 | 0.937 | 0.937 | 0.937 | 0.937 | 0.968 |
| δ | 4 | 4 | 4 | 4 | 4 | 4 | 2 |
| AC | 16 | 16 | 16 | 16 | 16 | 32 | 64 |
| σ | 8704 | 8704 | 8704 | 8704 | 8704 | 14464 | 16384 |
| AI | 3 | 3 | 3 | 3 | 3 | 3 | 3 |
| FP | 0 | 1.09 | 7 | 0 | 0 | 2 | 0 |
| OFP | 0 | 1.008 | 7 | 0 | 0 | 2 | 1 |
| SNR | 3.360 | 4.978 | 5.994 | 3.451 | 3.904 | 5.026 | 4.879 |
| TO | 5.694 | 5.740 | 5.801 | 5.734 | 5.694 | 5.734 | 5.730 |
| K | 0.109 | 0.262 | 0.664 | 0.622 | 0.454 | 0.216 | 0.238 |

Table 5: Comparing the evaluation properties of 7×7 s-boxes

| Properties | 7x7 | | | initial | proposed | WAGE | Niho |
|---|---|---|---|---|---|---|---|
| | min | avg | max | | | | |
| B | √ | √ | √ | √ | √ | √ | √ |
| NL | 54 | 54 | 54 | 54 | 54 | 54 | 54 |
| AD | 6 | 6 | 6 | 6 | 6 | 6 | 4 |
| CI | 0 | 0 | 0 | 0 | 0 | 0 | 0 |
| R | 0.984 | 0.984 | 0.984 | 0.984 | 0.984 | 0.937 | 0.984 |
| δ | 2 | 2 | 2 | 2 | 2 | 8 | 2 |
| AC | 24 | 24 | 24 | 24 | 24 | 80 | 16 |
| σ | 32768 | 32768 | 32768 | 32768 | 32768 | 86528 | 32768 |
| AI | 4 | 4 | 4 | 4 | 4 | 3 | 3 |
| FP | 0 | 0 | 0 | 0 | 0 | 0 | 2 |
| OFP | 0 | 0 | 0 | 0 | 0 | 4 | 1 |
| SNR | 4.806 | 6.670 | 8.077 | 5.318 | 5.84 | 6.628 | 7.062 |
| TO | 6.792 | 6.813 | 6.837 | 6.802 | 6.8 | 6.74 | 6.838 |
| K | 0.092 | 0.182 | 0.437 | 0.339 | 0.264 | 0.183 | 0.15 |

Construction of Side Channel Attacks Resistant S-boxes using Genetic Algorithms based on Coordinate Functions

Table 6: Comparing the evaluation properties of 8×8 s-boxes

| Properties | 8x8 | | | initial | **proposed** | Prince | Present |
|---|---|---|---|---|---|---|---|
| | min | avg | max | | | | |
| B | √ | √ | √ | √ | √ | √ | √ |
| NL | 112 | 112 | 112 | 112 | 112 | 112 | 112 |
| AD | 7 | 7 | 7 | 7 | 7 | 7 | 7 |
| CI | 0 | 0 | 0 | 0 | 0 | 0 | 0 |
| R | 0.984 | 0.984 | 0.984 | 0.984 | 0.984 | 0.984 | 0.984 |
| δ | 4 | 4 | 4 | 4 | 4 | 4 | 4 |
| AC | 32 | 32 | 32 | 32 | 32 | 32 | 32 |
| σ | 133120 | 133120 | 133120 | 133120 | 133120 | 133120 | 133120 |
| AI | 4 | 4 | 4 | 4 | 4 | 4 | 4 |
| FP | 0 | 1.062 | 6 | 0 | 0 | 0 | 0 |
| OFP | 0 | 1.192 | 6 | 0 | 0 | 0 | 0 |
| SNR | 8.253 | 9.541 | 10.76 | 9.599 | 8.758 | 9.694 | 9.323 |
| TO | 7.847 | 7.855 | 7.863 | 7.86 | 7.85 | 7.855 | 7.851 |
| K | 0.075 | 0.114 | 0.172 | 0.111 | 0.146 | 0.107 | 0.121 |

Table 7: Initial 4x4 s-box

| 8 | 0 | 1 | 10 | 9 | 4 | 2 | 6 |
| 11 | 7 | 14 | 12 | 5 | 15 | 13 | 3 |

Table 8: Our best generated 4x4 s-box

| 3 | 0 | 5 | 15 | 6 | 13 | 12 | 1 |
| 10 | 4 | 2 | 14 | 8 | 7 | 11 | 9 |

Table 9: Initial 5x5 s-box

| 8 | 0 | 26 | 17 | 22 | 28 | 29 | 24 |
| 19 | 16 | 4 | 6 | 7 | 18 | 16 | 23 |
| 13 | 31 | 25 | 30 | 2 | 20 | 12 | 1 |
| 5 | 3 | 15 | 27 | 9 | 21 | 10 | 11 |

Table 10: Our best generated 5x5 s-box

| 24 | 0 | 23 | 22 | 12 | 28 | 13 | 31 |
| 30 | 7 | 3 | 11 | 26 | 15 | 19 | 29 |
| 10 | 5 | 14 | 20 | 8 | 4 | 27 | 17 |
| 18 | 25 | 2 | 6 | 9 | 21 | 16 | 1 |

Table 11: Initial 6x6 s-box

| 22 | 25 | 37 | 10 | 14 | 5 | 60 | 15 |
| 0 | 7 | 26 | 63 | 50 | 59 | 48 | 23 |
| 6 | 62 | 24 | 38 | 16 | 58 | 32 | 61 |
| 43 | 20 | 29 | 4 | 52 | 33 | 35 | 12 |
| 13 | 56 | 44 | 54 | 51 | 47 | 42 | 27 |
| 28 | 40 | 11 | 55 | 9 | 36 | 41 | 45 |
| 46 | 8 | 31 | 34 | 17 | 2 | 18 | 19 |
| 30 | 39 | 49 | 1 | 3 | 57 | 21 | 53 |

Table 12: Our best generated 6x6 s-box

| 2 | 30 | 47 | 53 | 59 | 41 | 49 | 28 |
| 0 | 11 | 27 | 52 | 10 | 58 | 40 | 37 |
| 44 | 19 | 57 | 42 | 46 | 29 | 6 | 22 |
| 20 | 32 | 16 | 14 | 38 | 33 | 3 | 25 |
| 62 | 63 | 31 | 4 | 45 | 26 | 51 | 60 |
| 55 | 17 | 18 | 35 | 48 | 8 | 54 | 56 |
| 61 | 23 | 50 | 36 | 9 | 34 | 12 | 43 |
| 21 | 13 | 15 | 39 | 5 | 24 | 7 | 1 |

Table 13: Initial 7x7 s-box

| 66 | 114 | 56 | 86 | 115 | 85 | 11 | 78 |
| 124 | 71 | 44 | 3 | 41 | 87 | 4 | 81 |
| 104 | 10 | 34 | 15 | 108 | 48 | 2 | 16 |
| 95 | 92 | 65 | 67 | 55 | 62 | 28 | 97 |
| 76 | 57 | 12 | 96 | 6 | 18 | 120 | 91 |
| 54 | 35 | 79 | 100 | 109 | 69 | 121 | 9 |
| 36 | 126 | 111 | 77 | 74 | 45 | 125 | 122 |
| 73 | 90 | 26 | 98 | 58 | 80 | 51 | 72 |
| 118 | 33 | 21 | 116 | 123 | 117 | 82 | 61 |
| 110 | 64 | 68 | 99 | 19 | 37 | 46 | 53 |
| 83 | 31 | 50 | 24 | 93 | 89 | 52 | 60 |
| 84 | 22 | 63 | 107 | 25 | 38 | 88 | 7 |
| 39 | 30 | 17 | 13 | 119 | 102 | 8 | 14 |
| 5 | 29 | 20 | 127 | 70 | 106 | 43 | 112 |
| 59 | 49 | 101 | 1 | 47 | 113 | 32 | 0 |
| 23 | 94 | 40 | 75 | 27 | 103 | 105 | 42 |

Table 14: Our best generated 7x7 s-box

| 52 | 17 | 127 | 112 | 58 | 18 | 56 | 123 |
| 23 | 10 | 59 | 98 | 5 | 91 | 21 | 7 |
| 83 | 19 | 61 | 45 | 70 | 37 | 73 | 81 |
| 1 | 99 | 86 | 31 | 82 | 35 | 30 | 34 |
| 50 | 84 | 79 | 9 | 92 | 24 | 2 | 20 |
| 121 | 22 | 80 | 28 | 109 | 67 | 41 | 113 |
| 97 | 94 | 36 | 25 | 110 | 16 | 60 | 75 |
| 12 | 63 | 66 | 64 | 54 | 44 | 71 | 39 |
| 4 | 95 | 111 | 77 | 96 | 102 | 101 | 65 |
| 15 | 125 | 104 | 107 | 51 | 74 | 114 | 27 |
| 78 | 124 | 108 | 11 | 72 | 93 | 48 | 106 |
| 57 | 13 | 8 | 49 | 32 | 40 | 118 | 119 |
| 3 | 87 | 122 | 100 | 47 | 85 | 90 | 6 |
| 62 | 53 | 68 | 117 | 33 | 26 | 76 | 88 |
| 29 | 14 | 55 | 43 | 89 | 115 | 116 | 0 |
| 38 | 42 | 46 | 69 | 105 | 126 | 120 | 103 |

Table 14: Our best generated 7x7 s-box

| 52 | 17 | 127 | 112 | 58 | 18 | 56 | 123 |
|----|----|-----|-----|----|----|----|-----|
| 23 | 10 | 59 | 98 | 5 | 91 | 21 | 7 |
| 83 | 19 | 61 | 45 | 70 | 37 | 73 | 81 |
| 1 | 99 | 86 | 31 | 82 | 35 | 30 | 34 |
| 50 | 84 | 79 | 9 | 92 | 24 | 2 | 20 |
| 121 | 22 | 80 | 28 | 109 | 67 | 41 | 113 |
| 97 | 94 | 36 | 25 | 110 | 16 | 60 | 75 |
| 12 | 63 | 66 | 64 | 54 | 44 | 71 | 39 |
| 4 | 95 | 111 | 77 | 96 | 102 | 101 | 65 |
| 15 | 125 | 104 | 107 | 51 | 74 | 114 | 27 |
| 78 | 124 | 108 | 11 | 72 | 93 | 48 | 106 |
| 57 | 13 | 8 | 49 | 32 | 40 | 118 | 119 |
| 3 | 87 | 122 | 100 | 47 | 85 | 90 | 6 |
| 62 | 53 | 68 | 117 | 33 | 26 | 76 | 88 |
| 29 | 14 | 55 | 43 | 89 | 115 | 116 | 0 |
| 38 | 42 | 46 | 69 | 105 | 126 | 120 | 103 |

## Abbreviations

| | |
|---|---|
| AD | Algebraic Degree |
| AI | Algebraic Immunity, |
| CC | Confusion Coefficient |
| CI | Correlation Immunity |
| DPA | Differential Power Attack |
| DU | Differential Uniformity |
| FP | Fixed Point |
| NL | Nonlinearity Degree |
| OFP | Opposite Fixed Points |
| RDC | Robustness to Differential Cryptanalysis |
| S-Box | Substitution Box |
| SCA | Side Channel Attack |
| SNR | Signal-to-Noise Ratio |
| TO | Transparency Order |

## BIOGRAPHIES

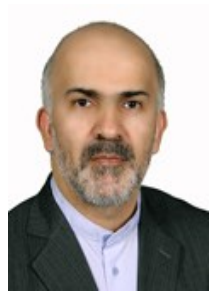

**Behrooz Khadem** (8 Aug. 1963) is an Associate Professor of the Imam Hossein comprehensive university in Tehran. He has graduated with applied mathematics B.Sc. in Tehran university (1991) and applied mathematics M.Sc. in Shaid Bahonar university in Kerman (1994) and applied mathematics (cryptography) Ph.D. in Kharazmi university in Tehran (2014). His interests are mathematical aspects of cryptography, data and protocol security, algorithms and artificial intelligence.

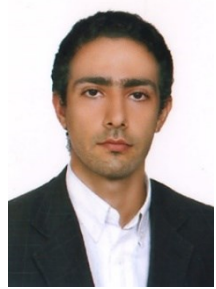

**Saeed Rajavzade** (14 April. 1989) is a Ph.D. student of the Payam Noor university in Tehran. He has graduated with mechanical engineering B.Sc. in Islamic Azad University of Tehran Markazi (2014) and pure mathematics M.Sc. in Maragheh University (2017). His interests are dynamical systems, nonlinear analysis, lattice and, mathematical aspects of cryptography, lattice-based cryptography, and Algorithms.

Table 15: Our best generated 8x8 s-box

| | | | | | | | | | | | | | | | |
|---|---|---|---|---|---|---|---|---|---|---|---|---|---|---|---|
| 62 | 248 | 28 | 48 | 229 | 103 | 173 | 102 | 33 | 116 | 149 | 194 | 97 | 147 | 228 | 134 | 62 |
| 109 | 223 | 110 | 27 | 25 | 70 | 120 | 208 | 11 | 245 | 58 | 209 | 73 | 211 | 212 | 183 | 109 |
| 184 | 203 | 138 | 187 | 166 | 179 | 195 | 135 | 159 | 142 | 240 | 78 | 186 | 141 | 84 | 254 | 184 |
| 90 | 178 | 23 | 136 | 54 | 29 | 61 | 133 | 51 | 76 | 193 | 231 | 9 | 246 | 232 | 225 | 90 |
| 252 | 4 | 155 | 44 | 31 | 177 | 17 | 77 | 143 | 94 | 217 | 131 | 46 | 96 | 121 | 190 | 252 |
| 251 | 151 | 130 | 170 | 216 | 64 | 26 | 56 | 243 | 214 | 249 | 146 | 160 | 0 | 52 | 224 | 251 |
| 156 | 127 | 148 | 132 | 128 | 201 | 181 | 60 | 81 | 2 | 47 | 20 | 124 | 85 | 105 | 153 | 156 |
| 80 | 111 | 157 | 244 | 63 | 67 | 83 | 137 | 71 | 117 | 182 | 32 | 139 | 112 | 235 | 41 | 80 |
| 114 | 74 | 219 | 3 | 57 | 45 | 10 | 140 | 113 | 145 | 108 | 18 | 13 | 144 | 232 | 50 | 114 |
| 238 | 119 | 191 | 162 | 21 | 104 | 196 | 88 | 14 | 233 | 72 | 91 | 107 | 230 | 176 | 226 | 238 |
| 168 | 30 | 165 | 68 | 43 | 125 | 253 | 164 | 118 | 115 | 40 | 206 | 218 | 188 | 175 | 255 | 168 |
| 19 | 180 | 204 | 174 | 37 | 234 | 172 | 16 | 49 | 75 | 213 | 66 | 189 | 227 | 126 | 122 | 19 |
| 199 | 106 | 36 | 7 | 98 | 247 | 198 | 236 | 69 | 154 | 167 | 6 | 169 | 222 | 5 | 161 | 199 |
| 210 | 171 | 192 | 53 | 197 | 1 | 42 | 79 | 59 | 35 | 100 | 99 | 202 | 55 | 93 | 22 | 210 |
| 15 | 129 | 200 | 86 | 65 | 152 | 207 | 158 | 8 | 38 | 39 | 92 | 89 | 250 | 24 | 82 | 15 |
| 220 | 12 | 101 | 123 | 150 | 205 | 87 | 221 | 241 | 185 | 215 | 237 | 95 | 163 | 34 | 239 | 220 |